\documentstyle[aps,prb]{revtex}
\tightenlines
\begin{document}

\title{Ion Sizes and Finite-Size Corrections for Ionic-Solvation Free
  Energies}

\author{Gerhard Hummer, Lawrence R. Pratt and Angel E. Garc\'{\i}a}

\address{Theoretical Division, MS K710, Los Alamos National Laboratory,
Los~Alamos, New Mexico 87545\\ Phone: (505) 665-1923; Fax: (505) 665-3493;
e-mail: hummer@lanl.gov}

\date{in press: {\em J. Chem. Phys.; LA-UR 97-1591}}

\maketitle

\begin{abstract}
  Free energies of ionic solvation calculated from
  computer simulations exhibit a strong system size dependence.  We
  perform a finite-size analysis based on a dielectric-continuum model
  with periodic boundary conditions.  That analysis results in an
  estimate of the Born ion size.  Remarkably, the finite-size
  correction applies to systems with only eight water molecules
  hydrating a sodium ion and results in an estimate of the Born radius
  of sodium that agrees with the experimental value.
\end{abstract}

Calculation of ionic-hydration free energies from computer simulations
require great care to avoid artifacts due to the long-range
electrostatic interactions.  We have recently shown that Ewald
summation\cite{Ewald:21,deLeeuw:80:a} can give results for single-ion
free energies that are essentially system-size independent for as few
as about 16 water molecules.\cite{Hummer:96:a} That has been achieved
by including the self-interactions $0.5 q^2 \xi_{\rm Ew}$ of the ion
($\xi_{\rm Ew}\approx -2.837297/L$ with $L$ the length of the cubic
box).  Ewald summation determines the electrostatic interactions using
lattice sums for a periodically replicated simulation box.  This
provides a natural description of the electrostatics in the periodic
space resulting from the periodic boundary conditions commonly used in
computer simulations.  Deviations from the approximate finite-size
correction\cite{Hummer:96:a,Hummer:95:e} are expected for solvents
with finite dielectric constant $\epsilon<\infty$ (i.e., in the
infinite-dilution limit of an ion in a non-conducting
solvent),\cite{Figueirido:97,Lynden-Bell:97} and if the ion size is
comparable to the dimensions of the simulation box.

The self-interaction of an ion is introduced in the Ewald summation
through interactions with the compensating charge background implicit
in the Ewald summation approach and with the periodic images of the
ion.  A similar correction has been developed for free energies of
polar molecules.\cite{Hummer:95:e} Figueirido et
al.\cite{Figueirido:97} used a point ion in a homogeneous dielectric
medium as a model to extend the system-size corrections to solvents
with finite dielectric constants $\epsilon$, finding that $0.5 q^2
\xi_{\rm Ew} (1-1/\epsilon)$ should be a good approximation to the
finite-size effects.  Here, we further extend Figueirido et al.'s
study\cite{Figueirido:97} to ions of finite size.  We show that the
ion-size correction is important for small system sizes.  We also show
that the analysis of the system-size dependence of ionic free energies
leads to the definition of an ion size in dielectric media in
excellent agreement with the Born radius.\cite{Born:20}

We approximate the free energy of charging an ion as the sum of the
explicit simulation contribution $F_{\rm sim}$ and a self-interaction
$0.5 q^2\xi_{\rm Ew}$, as in Ref.~\onlinecite{Hummer:96:a}.  To
account for further effects of finite system size, we add the
difference of charging an ion in an infinite and finite system,
$F_{\infty}-F_L$,
\begin{eqnarray}
  \label{eq:free}
  F_0 & = & F_{\rm sim} + \frac{1}{2}q^2\xi_{\rm Ew} + \left( F_\infty -
  F_L \right)
\end{eqnarray}
The finite-size correction $F_{\infty}-F_L$ is evaluated for a
simplified model of a point charge $q$ in a spherical cavity with
radius $R$ inside a dielectric continuum with dielectric constant
$\epsilon$, as schematically shown in Fig.~\ref{fig:scheme}.  The
finite system is formed by an ion in a periodically replicated box of
length $L$, where the box is charge neutral through addition of a
homogeneous background charge with density $-q/L^3$.  The infinite
system is obtained by taking the limit $L\rightarrow\infty$ and
corresponds to the Born model.\cite{Born:20} If the model used to
calculate $F_\infty-F_L$ were exact, $F_L$ would cancel $F_{\rm
  sim}+0.5 q^2\xi_{\rm Ew}$.  However, the dielectric model is only an
approximation to the simulation system with explicit solvent, making
the finite-size correction only approximate.

The finite-size correction is the difference of charging the ion in an
infinite ($L\rightarrow\infty$) and finite system,
\begin{eqnarray}
  \label{eq:limit}
  F_\infty - F_L & = & \frac{1}{2} q^2 \left[ 
    \xi(\epsilon,R,L\rightarrow\infty) - \xi(\epsilon,R,L) \right]~,
\end{eqnarray}
$\xi(\epsilon,R,L)$ is the electrostatic potential minus the bare
self-interaction at the position of a unit charge,
\begin{eqnarray}
  \label{eq:self}
  \xi(\epsilon,R,L)=\lim_{r\rightarrow 0} \left[\phi({\bf r})-1/r\right]~.  
\end{eqnarray}
We determine the electrostatic potential $\phi({\bf r})$ for the
dielectric model described above by solving the corresponding Poisson
equation,
\begin{eqnarray}
  \label{eq:Poisson}
  \nabla \left[ \epsilon({\bf r}) \nabla \phi({\bf r}) \right] & = &
  - 4 \pi \left[ \delta({\bf r}) - 1 / L^3 \right].
\end{eqnarray}
We define potentials $\phi_-({\bf r})$ and $\phi_+({\bf r})$ inside
and outside the sphere.  For $R<L/2$, the boundary conditions are
\begin{mathletters}
  \label{eq:boundary}
  \begin{eqnarray}
    \label{eq:boundary1}
    \phi_-({\bf r}) & = & \phi_+({\bf r}) \quad \mbox{for } |{\bf r}|=R~,\\
    \label{eq:boundary2}
    \frac{\partial\phi_-({\bf r})}{\partial r} & = &
    \epsilon\frac{\partial\phi_+({\bf r})}{\partial r} \quad \mbox{for
    }    |{\bf r}|=R~,\\
    \label{eq:boundary3}
    \frac{\partial\phi_+({\bf r})}{\partial x} & = & 0 \quad
    \mbox{for } |x|=L/2~.
\end{eqnarray}
\end{mathletters}
The last condition reflects the periodicity of the potential and
applies analogously for $y$ and $z$.

To solve this electrostatic problem, we expand $\phi_-$ and $\phi_+$
into a complete set of functions that satisfy Poisson's equation
Eq.~(\ref{eq:Poisson}) and then choose the expansion coefficients to
satisfy the boundary conditions Eq.~(\ref{eq:boundary}).  Such
functions are the kubic-harmonic polynomials $K_n$ introduced by von
der Lage and Bethe.\cite{vdLage:47}  We define
\begin{mathletters}
  \label{eq:expansion}
  \begin{eqnarray}
    \label{eq:expansion1}
    \phi_-({\bf r}) & = & \frac{1}{r}+\frac{2\pi r^2}{3L^3} +
    \sum_{n=2}^{N/2} a_{2n} K_{2n}({\bf r}) + C~,\\
    \label{eq:expansion2}
    \phi_+({\bf r}) & = & \frac{b_{-1}}{r}+b_0+\frac{2\pi
      r^2}{3\epsilon L^3} + \sum_{n=2}^{N/2} b_{2n} K_{2n}({\bf r}) + C~.
  \end{eqnarray}
\end{mathletters}
For a finite set of functions $K_{2n}$ up to polynomial order $2n\leq
N$, the solution is only approximate.  We choose the coefficients in
the least-square sense, defining a $\chi^2$ functional,
\begin{eqnarray}
  \label{chi-square}
  \chi^2 & = & R^2\int_0^{2\pi} d\varphi \int_0^\pi \sin\theta\;
    d\theta \Biggl\{ \left[ \phi_-({\bf r}) - \phi_+({\bf r})\right]^2
    \nonumber\\
    && \left. + L \left[ \frac{\partial\phi_-({\bf r})}{\partial r} -
        \epsilon\frac{\partial\phi_+({\bf r})}{\partial r}\right]^2
    \right\}_{|{\bf r}|=R}\nonumber\\ &&
    + 6{\int\int}_{y^2+z^2 < L^2/4} dy\;dz \left[
    \frac{\partial\phi_+({\bf r})}{\partial x}\right]^2_{x=L/2}~,
\end{eqnarray}
with spherical polar coordinates $r$, $\varphi$ and $\theta$.  The
last integral extends only over a circular region because this results
in an analytical solution of that integral.  Also, the deviations from
Eq.~(\ref{eq:boundary3}) of truncated kubic-harmonic expansions of the
Ewald potential\cite{Hummer:95:e,Slattery:80,Adams:87,Hummer:93} are
largest in the corners of the box.  Fitting only a circular region
therefore produces smaller distortions of the potential at the center
of the box.

Minimizing $\chi^2$ with respect to the coefficients $b_{-1}$, $b_0$,
$a_4$, $b_4$ etc. results in a set of linear equations for those
coefficients.  Using the computational algebra package
REDUCE,\cite{REDUCE} we solve for the unknown expansion coefficients
for different orders $N$ of the kubic-harmonic expansion.  We then
apply Taylor expansion with respect to $1/\epsilon$ and the size
parameter $R/L$.  This results in an asymptotic expansion of
$\xi(\epsilon,R,L)$ as
\begin{eqnarray}
  \label{eq:asymptotic}
  \xi(\epsilon,R,L) & = & \frac{c_0}{\epsilon L} - \frac{2\pi R^2}{3L^3}
  \frac{\epsilon-1}{\epsilon} + \frac{c_2 R^2}{L^3\epsilon^3}+
  \frac{c_4 R^4}{L^5\epsilon^3}
  \nonumber\\  &&
  + {\cal O}\left(\frac{R^5}{L^6},\epsilon^{-5}\right)~.
\end{eqnarray}
Notice that because of the finite system size, $\phi({\bf r})$ is
normalized by adding a constant $C$ such that the average potential in
the simulation box is
zero,\cite{Hummer:96:a,Hummer:93,Figueirido:95,Brush:66,Nijboer:88}
\begin{eqnarray}
  \label{eq:normalize}
  \int_{\rm box} d{\bf r} \; \phi({\bf r}) & = & 0~,
\end{eqnarray}
rather than $\phi({\bf r})\rightarrow 0$ for $r\rightarrow\infty$.
The familiar Born term $0.5(1-\epsilon^{-1})/R$ is thus contained in
the normalization constant $C$.

The coefficients $c_0$, $c_2$ and
$c_4$ are listed in Table~\ref{tab:xi}.  As the order $N$ of expansion
functions increases, the coefficient $c_0$ converges towards the
self-energy of a point charge in a cubic Wigner
lattice,\cite{Nijboer:88} $c_0 \rightarrow L\xi_{\rm Ew} \approx
-2.837297$.  The coefficients $c_2$ and $c_4$ appear to converge to
values close to zero.  The most interesting term in
Eq.~(\ref{eq:asymptotic}) is the lowest order correction for the
finite size of an ion, $-2\pi(\epsilon-1) R^2 / (3\epsilon L^3)$.
Eq.~(\ref{eq:asymptotic}) converges to the correct limit for
$\epsilon\rightarrow\infty$, which can be found independently using
the continuity of $\phi({\bf r})$ at the dielectric interface $|{\bf
  r}|=R$ and the spherical geometry for $\epsilon\rightarrow\infty$,
\begin{eqnarray}
  \label{eq:infinity}
  \xi(\epsilon\rightarrow\infty,R,L)&=&
  - \frac{2\pi R^2}{3L^3} + \frac{16\pi^2 R^5}{45 L^6}~.
\end{eqnarray}
The last term is contained in the ${\cal O} \left(
  R^5/L^6,\epsilon^{-5} \right)$ term of Eq.~(\ref{eq:asymptotic}).

We can now calculate the finite-size correction to the free energy of
an ion with radius $R$ in a dielectric medium with Ewald-summation
electrostatics,
\begin{eqnarray}
  \label{eq:fs}
  F_\infty - F_L = -\frac{1}{2} q^2 \xi(\epsilon,R,L)~,
\end{eqnarray}
using Eq.~(\ref{eq:limit}) and $\xi(\epsilon,R,L\rightarrow\infty)=0$
for the particular choice of $C$ [Eq.~(\ref{eq:normalize})].  This
results in an approximation to the free energy $F_0$ of charging the
ion including finite-size effects:
\begin{eqnarray}
  \label{eq:finite-size}
  F_0 & \approx & F_{\rm sim} + \frac{1}{2} q^2
  \frac{\epsilon-1}{\epsilon} \left
  (\xi_{\rm Ew}+\frac{2\pi R^2}{3L^3}\right)~.
\end{eqnarray}
In the limit $\epsilon\rightarrow\infty$, the finite-size correction
$F_\infty-F_L$ does not contain terms of order $L^{-1}$ and $L^{-2}$.
This explains the success of using $F_{\rm sim} + 0.5 q^2 \xi_{\rm
  Ew}$ alone for the free energy of charging an ion in a
conducting\cite{Hummer:93} or highly polar\cite{Hummer:96:a}
environment, without further finite-size correction that takes the ion
size or the dielectric constant of the solvent into account.

The validity of the approximate finite-size correction
Eq.~(\ref{eq:finite-size}) is illustrated in Fig.~\ref{fig:correction}
for a sodium ion in water.  We fit the sodium-ion data $F_{\rm sim}$
for electrostatic free energies calculated from simulations with $M=8$
to $M=256$ water molecules\cite{Hummer:96:a} to
Eq.~(\ref{eq:finite-size}) with $F_0$ and $R$ as parameters.  For the
dielectric constant, $\epsilon\rightarrow\infty$ is used, but values
of $\epsilon\approx 80$ have little effect on the result.  That fit
yields $F_0=-402.5\pm 1$ kJ mol$^{-1}$ for the electrostatic
contribution to the solvation free energy of sodium.  The radius $R$
of the sodium ion is found to be $R=0.18$ nm, in excellent agreement
with the effective Born radius $0.180$ nm of that ion, as determined
by Latimer, Pitzer and Slansky\cite{Latimer:39} or, more recently,
0.188~nm, as compiled by Marcus.\cite{Marcus:91:a} In the fit, the ion
radius $R$ is determined mostly by the data for small system sizes.
Using only the $M=8$ and 16 data and fixing $F_0$ at $-402.5$ kJ
mol$^{-1}$, a fit of $R$ yields 0.159 nm ($M=16$), 0.190 nm ($M=8$)
and 0.184 nm ($M=8$ and 16).

The simulation result for the Born radius is
0.172~nm.\cite{Hummer:96:a} That is, $F_0$ and $R$ are self-consistent
assuming an uncertainty greater than 0.01 nm in $R$.  We can therefore
fit the free-energy data with the solvation free energy $F_0$ as the
only parameter by substituting the Born expression $F_0\approx -0.5
q^2 (1-\epsilon^{-1})/R$ for $R$,
\begin{eqnarray}
  \label{eq:fit}
  F_0 & \approx & F_{\rm sim} + \frac{1}{2} q^2
  \frac{\epsilon-1}{\epsilon}
  \left(\xi_{\rm Ew}+\frac{\pi q^4 (1-\epsilon^{-1})^2} {6 {F_0}^2
  L^3}\right)~.
\end{eqnarray}
That fit yields $F_0=-403\pm 1$ kJ mol$^{-1}$ (for
$\epsilon\rightarrow\infty$).  These results show that the
system-size effects on the hydration of sodium can be described
accurately for $M\geq 8$ water molecules by the dielectric continuum
model with periodic boundary conditions shown in
Fig.~\ref{fig:scheme}.  It is remarkable that continuum-electrostatics
approximations apply even for as few as eight water molecules
solvating the sodium ion provided the periodic boundary conditions are
considered.  Of course, determination of the value of $F_0$ and the
Born radius requires further molecular considerations.  To treat
solutes with more complicated shapes will require additional
computational effort, as was suggested previously.\cite{Pratt:97}

In summary, we have found an approximate finite-size correction for
ions in water that takes into account the size of the ions as well as
the dielectric constant of the solvent.  These findings extend the
previous work by Hummer et al.\cite{Hummer:96:a,Hummer:95:e,Hummer:93}
and Figueirido et al.\cite{Figueirido:97,Figueirido:95} The results of
Refs.~\onlinecite{Hummer:96:a} and \onlinecite{Hummer:93} are
recovered in the limit of a conducting solvent
($\epsilon\rightarrow\infty$) and a point ion $R/L\rightarrow 0$.  In
the limit of a point ion but for a finite dielectric constant
$\epsilon<\infty$, we obtain the finite-size correction of
Ref.~\onlinecite{Figueirido:97}.  For strongly polar solvents, such as
water ($\epsilon\approx 80$), and typical system sizes of hundred or
more solvent molecules with ions of comparable size, the difference to
the finite-size correction obtained
previously\cite{Hummer:96:a,Hummer:95:e} for
$\epsilon\rightarrow\infty$ and $R/L\rightarrow 0$ will be small
($\lesssim$ 1-2\%).  We have found that the finite-size correction
derived in this paper gives qualitatively and quantitatively correct
behavior. This finite-size analysis results in an estimate of the
sodium-ion size that is in agreement with the experimental data.

\acknowledgments We want to thank Dr. F. Figueirido for sending us a
preprint of Ref.~\onlinecite{Figueirido:97}.  G.H. wants to thank
Prof.  M.  Neumann for valuable discussions about solving
electrostatic problems in periodic systems, and Dr. R. LaViolette for
discussions about kubic-harmonic polynomials, specifically about using
normal derivatives at the boundary to fit coefficients.

\begin{table}
  \caption{Coefficients of the approximation
    Eq.~(\protect\ref{eq:asymptotic}) to $\xi$ for different levels
    $N$ of kubic-harmonic expansions.
    ($N=2$ means that no kubic harmonics have been used.)\label{tab:xi}}
  \begin{tabular}{rrrr}
    $N$ & $c_0$ & $c_2$ & $c_4$ \\ \hline
    2  & $-2.9037$ & $ 0.9751$ & $-0.8581$ \\
    4  & $-2.8309$ & $-0.1638$ & $ 0.1558$ \\
    6  & $-2.8362$ & $-0.0728$ & $ 0.0711$ \\
    8  & $-2.8398$ & $ 0.0371$ & $-0.0366$ \\
    10  & $-2.8370$ & $ 0.0022$ & $-0.0023$ \\
  \end{tabular}
\end{table}

\begin{figure}[htbp]
  \caption{Schematic representation of the contributions to the
    electrostatic solvation free energy of an ion, $F_{\rm sim} +
    0.5 q^2\xi_{\rm Ew} + ( F_\infty - F_L )$.}
  \label{fig:scheme}
\end{figure}

\begin{figure}[htbp]
  \caption{Finite-size correction for the hydration free energy of a
    sodium ion.  Shown is the free energy of charging a sodium ion
    in water from charge zero to $e$ as a function of the inverse
    simulation box length $1/L$.  The symbols are simulation data
    for Ewald-summation electrostatics from
    Ref.~\protect\onlinecite{Hummer:96:a} for $M=8$, 16, 32, 64,
    128, and 256 water molecules.  The dashed line is the fit to the
    correction formula of Ref.~\protect\onlinecite{Hummer:96:a} (for
    $M\geq 16$), corresponding to Eq.~(\protect\ref{eq:finite-size})
    with $\epsilon\rightarrow\infty$ and $R=0$.  That fit yields an
    extrapolated free energy $F_0$ of $-405$ kJ mol$^{-1}$. The
    solid line shows the fit to the correction formula
    Eq.~(\protect\ref{eq:finite-size}) with
    $\epsilon\rightarrow\infty$, where the sodium-ion radius $R$ and
    the free energy of charging $F_0$ are estimated to be $0.18$ nm and
    $-402.5$ kJ mol$^{-1}$, respectively.}
  \label{fig:correction}
\end{figure}

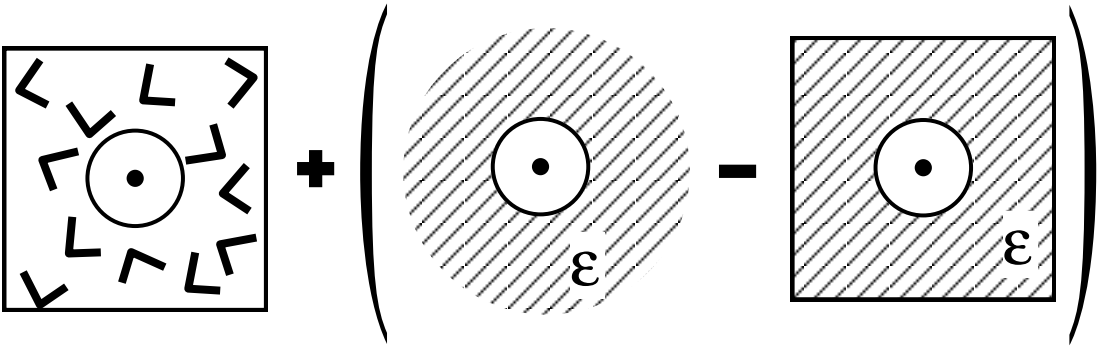
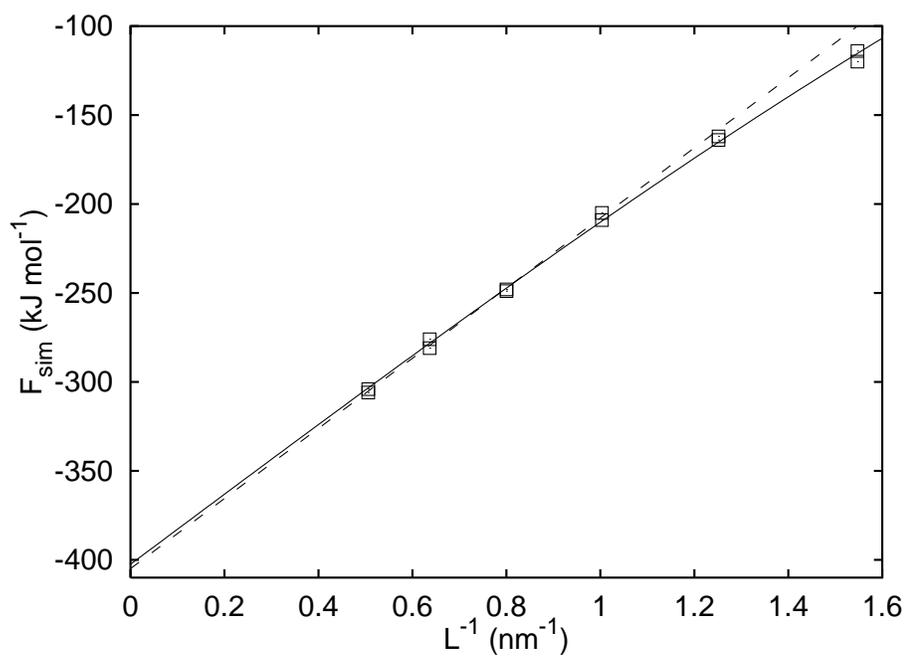
\end{document}